\documentclass{elsart}
\usepackage{amsmath}
\usepackage{amssymb}
\usepackage{graphics}
\begin{document}

\begin{frontmatter}
\title{Crossing symmetry and phenomenological widths in effective Lagrangian
models of the pion photoproduction process}
\author[UCM]{C. Fern\'andez-Ram\'{\i}rez\corauthref{cor}},
\corauth[cor]{Present address: Center for Theoretical Physics, 
Massachusetts Institute of Technology.
77 Massachusetts Ave., Cambridge, MA 02139-4307, USA.}
\ead{cesar@nuc2.fis.ucm.es}
\author[UCM,IEM]{E. Moya de Guerra},
\author[UCM]{J.M. Ud\'{\i}as}
\address[UCM]{Grupo de F\'{\i}sica Nuclear, 
Departamento de F\'{\i}sica At\'omica, Molecular y Nuclear, 
Facultad de Ciencias F\'{\i}sicas, 
Universidad Complutense de Madrid,
Avda. Complutense s/n, E-28040 Madrid, Spain}
\address[IEM]{Instituto de Estructura de la Materia, CSIC,
Serrano 123, E-28006 Madrid, Spain}
\begin{abstract}
We investigate the importance of crossing symmetry in effective field 
models and the effects of phenomenological nucleon resonance widths
on the paradigmatic case of pion photoproduction.
We use reaction models containing four star resonances up to 1.8 Gev
($\Delta$(1232), N(1440), N(1520), N(1535), $\Delta$(1620), N(1650),
$\Delta$(1700), and N(1720)) with different prescriptions for
crossed terms and widths, to fit the latest world database
on pion photoproduction. We compare $\chi^2$ results from selected multipoles and fits.
The $\chi^2$ is highly dependent on the fulfillment of  crossing symmetry and the
inclusion of $u$ channels.
\end{abstract}
\begin{keyword}
Crossing symmetry \sep phenomenological width \sep pion photoproduction
\PACS  13.60.Le \sep 25.20.Lj
\end{keyword}
\end{frontmatter}

In the non-perturbative regime of Quantum Chromodynamics (QCD)
we have to rely on effective field models to describe physical processes
governed by the strong interaction. This is particularly so
in the energy region of the nucleon mass and its excitations.
In an effective field model we build suitable Lagrangians to describe
particle couplings compatible with the symmetries of the underlying 
fundamental theory (QCD)
and we use a perturbative approach to calculate
the physical observables.
In this scheme, the reliability of any reaction model
counts on the soundness of the model
framework and on the fulfillment of the symmetries of the 
underlying theory.
Following this reasoning, the reliability of a
complex calculation on nuclei starting from an
elementary reaction model (e.g., meson exchange currents \cite{MEC} 
or pion photoproduction from nuclei
starting from a model on pion photoproduction from the nucleon \cite{FMVU07}) 
relies on how sound is the theoretical background used in the construction
of the elementary reaction model.

At tree level, the invariant amplitudes we obtain from the effective 
Lagrangians are real. 
Hence, unitarity of the scattering matrix is not respected
but, as long as we include all the Feynman diagrams emerging from the 
effective theory, crossing symmetry is fulfilled.
In a perturbative effective field theory
it is assumed that unitarity should be restored
once we include the higher order effects.
The exact calculation of higher orders is an overwhelmingly complex task,
so it is customary in the development of reaction models 
\cite{fernandez06a,fernandez06b,Drechsel,Benmerrouche91,Garcilazo93,Fer07}
to take into account higher order terms effectively through form factors, 
final and initial state interactions (FSI and ISI),
and phenomenological widths 
(dressing of the propagators). In doing so, one often pays 
the price of breaking symmetries that 
the theory should respect, such as crossing symmetry and consistency
between widths and couplings.
Since unitarity and crossing symmetry are linked together 
fixed-$t$ dispersion relations are also another method to satisfy these symmetries
\cite{DispersionRelations}.

Among all the required symmetries, crossing is a symmetry that 
the invariant amplitude of any well-defined effective field
theory has to fulfill \cite{Weinberg79}.
Similarly to what happens with gauge invariance \cite{fernandez06a,Nozawa90},
there are different options to implement crossing symmetry.
In this case, one of the options is to start from an
$A^{\text{bare}}$ which breaks crossing symmetry, and 
to build an FSI amplitude
($A^{\text{FSI}}$) that restores the symmetry, so that the
complete $A$ amplitude respects crossing symmetry. 
The other option is to build diagrams that are explicitly crossing symmetric:
$A^{\text{bare}}$ is crossing symmetric and $A^{\text{FSI}}$ does not break
this symmetry.
In this paper we employ this second scheme to the case of 
photo pion production from free nucleons. 
FSI can be taken into
account as a distortion of the pion wave function 
\cite{fernandez06a,fernandez06b,Watson}.
$A^{\text{bare}}$ is just the tree-level Feynman diagram but it includes 
the width in the propagator 
and corresponding form factor in the hadronic vertex. 
We split
the invariant amplitude in two parts: a background given by
the Born and vector meson exchange terms, and the contribution
of the nucleon resonances. 
In what follows we focus on
the contribution of nucleon resonances to $A^{\text{bare}}$.
We discuss the effect of the crossed terms using different
types of widths and different prescriptions on the resonance propagators and
amplitudes.
We have studied typical non-resonant multipoles because of their background character, what relates them to $u$ channel contributions.

Focusing first on the resonant contribution to the invariant amplitude,
we recall that, at tree level, both
the direct and the crossed contributions to the invariant amplitude
are real.
The inclusion of
a phenomenological width changes the scenario providing
the amplitude with an imaginary part.
When one considers direct and crossed terms one usually includes
a resonance width in the $s$ channel but not in the $u$ channel
\cite{Benmerrouche91,Garcilazo93}:
\begin{equation}
A^{\text{res}}\left( s,u \right)
=h\frac{A(s)}{s-M^{*2}+iM^*\Gamma \left( s \right)}
+h\frac{B(u)}{u-M^{*2}} . \label{eq:garcilazo}
\end{equation}

In what follows we refer to the choice of  
Eq. (\ref{eq:garcilazo}) as model V, for reasons that will become apparent.

The width is included in the term $iM^*\Gamma$ in the denominator,
compatible with what is obtained dressing the propagator with
pions \cite{dressing}. $h$ stands for the strong coupling constant and $M^*$
for the mass of the resonance. 
The width $\Gamma$ is defined as
\begin{equation}
\Gamma \left(s \right) 
= \sum_j \Gamma_j X_j \left( s \right)
\end{equation}
where $j = \pi , \pi \pi , \eta$ stands for the different decay
channels, $X_j \left( s \right)$ accounts for
the energy dependence of the width, and $\Gamma_\pi \propto h^2$.

Model V 
breaks crossing symmetry. One may think that this is not the case,
because taking a zero width in the crossed channel is equivalent
to include a width $\Gamma=\Gamma (u)$ in the $u$ channel 
($\Gamma (u)=0$ as $u<0$).
Although crossing symmetry may seem formally respected when one includes
$\Gamma=\Gamma (u)$ in the $u$ channel there is an inconsistency
between strong couplings and widths.
Consistency requires that the energy dependence that appears in the width 
is taken into account in the strong vertex. This means that the direct
and the crossed terms should contain the form factors $\sqrt{X_\pi(s)}$ and
$\sqrt{X_\pi(u)}$ respectively.
Hence, a zero width in the crossed channel would imply a null
contribution from the $u$ channel. 
From this point of view, only the two forthcoming 
choices remain consistent.
\begin{enumerate}
\item[(i)] One choice is to include
the energy dependence of the width as 
a form factor in the amplitude removing completely the $u$ channel,
\begin{equation}
A^{\text{res}}\left( s,u \right)
=h\sqrt{X_\pi \left(s \right)}
\frac{A(s)}{s-M^{*2}+iM^*\Gamma \left( s \right)}+0 \label{eq:width1}
\end{equation}

We call to this choice, Eq. (\ref{eq:width1}), model IV.
In the limiting case where the width is a constant
($\Gamma (s)= \Gamma_0$; $X_\pi \left( s \right)=1$)
Eq. (\ref{eq:width1}) transforms into
\begin{equation}
A^{\text{res}}\left( s,u \right)
=h\frac{A(s)}{s-M^{*2}+iM^*\Gamma_0} + 0  \label{eq:width3},
\end{equation}
which we call model III.

From an effective field theory point of view the complete disappearance
of the $u$ channel does not seem sensible but it cannot 
\textit{a priori} be discarded.

\item[(ii)] The other consistent choice is to include an energy-dependent width 
which depends on both $s$ and $u$ Mandelstam variables
and contributes to both direct and crossed terms \cite{fernandez06a}:
\begin{equation}
\begin{split}
A^{\text{res}}\left( s,u \right)&=
h\sqrt{X_\pi \left(s,u\right)}
\frac{A(s)}{s-M^{*2}+iM^*\Gamma \left( s,u \right)} \\
&+h\sqrt{X_\pi \left(s,u\right)}
\frac{B(u)}{u-M^{*2}+iM^*\Gamma \left( s,u \right)}.
\end{split} \label{eq:width2}
\end{equation}

The width  $\Gamma(s,u)$ in Eq. (\ref{eq:width2}) is defined as
\begin{equation}
\Gamma \left(s,u \right) 
= \sum_j \Gamma_j X_j \left( s , u \right) ,\label{eq:width}
\end{equation}
with
\begin{equation}
X_j \left( s , u \right) \equiv X_j \left( s \right) +  X_j 
\left( u \right) - X_j \left( s \right)  X_j \left( u \right).
\label{eq:Xj}
\end{equation}

This is the choice that we took in Refs. 
\cite{fernandez06a,fernandez06b,fernandez06c} and that we call here
model I.

In the limiting case where the width is a constant ($\Gamma_0$) we get
from (\ref{eq:width2}):
\begin{equation}
A^{\text{res}}\left( s,u \right)
=h\frac{A(s)}{s-M^{*2}+iM^*\Gamma_0}
+h\frac{B(u)}{u-M^{*2}+iM^*\Gamma_0}, \label{eq:width4}
\end{equation}
which we call model II.
\end{enumerate}

\begin{figure}
\begin{center}
\rotatebox{0}{\scalebox{0.4}[0.4]{\includegraphics{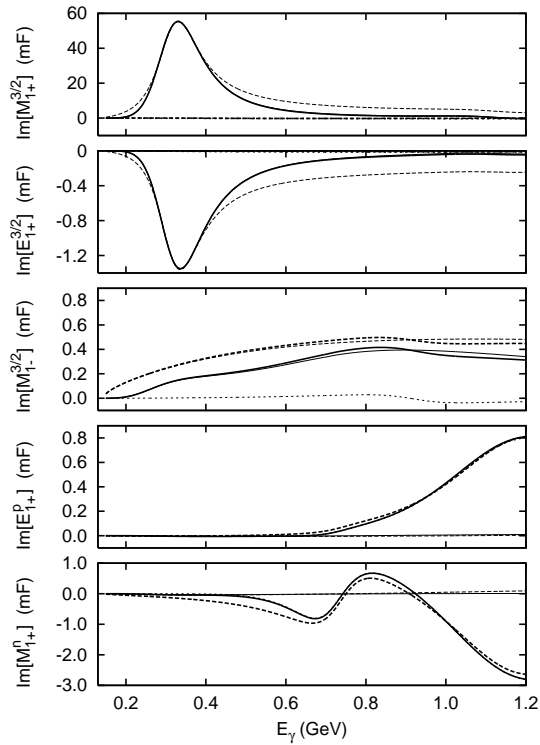}}}
\end{center}
\caption{Contribution of $s$ and $u$ channels to
bare electromagnetic multipoles. Curve conventions: 
Thick solid: Complete calculation with energy-dependent width model I;
Thin solid: Contribution of $u$ channel to model I;
Thick dashed: Complete calculation with a constant width model II;
Thin dashed: Contribution of $u$ channel to model II;
Short dashed: Contribution of $s$ channel to model I. 
All the curves have been obtained with the same set of coupling constants, that is 
the ones obtained fitting data with model I.}
\label{fig:fig-crossing}
\end{figure}

With choice (ii), the $u$ channel also contributes to the imaginary part
of the electromagnetic multipoles.
However, the imaginary part of the $u$ channel contributes differently
to the multipole amplitudes than the direct term,
acting as a background. This can be seen in Fig. \ref{fig:fig-crossing}
where we show the $u$ channel contributions to models I and II.
We have analyzed these contributions to every multipole.

In Fig. \ref{fig:fig-crossing} we show examples, extracted from our
extensive analysis, of the imaginary parts of five
bare electromagnetic multipoles for pion photoproduction
using the same parameter set for models I and II.
The parameter set used is the one obtained by fitting data with model I.
We focus on the imaginary part of the multipoles because it is there
where the contribution of the phenomenological widths shows up more clearly.
The contribution of Born and meson-exchange terms to the bare amplitudes
is real. Hence, the imaginary part of the bare multipoles shown in
Fig. \ref{fig:fig-crossing} comes only from 
the $s$ and the $u$ channels of the resonances.
We can see that the effect of a width in the $u$ channel  
in the multipoles $E_{1+}^p$ and $M_{1+}^n$ is very small,
independently on whether the width is energy-dependent or constant. In this article we use the 
standard notation for the electromagnetic multipoles in pion photoproduction \cite{SAID}:
$E$ and $M$ stand for electric and magnetic multipoles; the superindex stands for the
isospin: 3/2 for isospin-3/2, $p$ for proton isospin-1/2, and $n$ for neutron isospin-1/2;
the first subindex is related to the relative orbital angular momentum ($L$) of the outgoing pion and nucleon: it takes the value $L$ for the electric multipole and $L+1$ for magnetic one; and the second subindex stands for the parity: $+$ or $-$.
The same behavior is observed for all the multipoles with the exception of
the $M_{1-}^{3/2}$ multipole. The $M_{1-}^{3/2}$ 
is dominated by the $u$ channel and shows
a stronger dependence on the character of the width
(constant or energy-dependent).
The $s$ channel contribution to this multipole (short-dashed curve)
comes basically from the $\Delta(1620)$ resonance .
For completeness we also provide results on the imaginary part of the two resonant multipoles
of the $\Delta(1232)$ ($M^{3/2}_{1+}$ and $E^{3/2}_{1+}$) where it can be seen that the $u$ channel
contribution to the $M^{3/2}_{1+}$ multipole is zero as expected.

From this analysis 
we may conclude that for most of the multipoles the overall behavior
can be obtained with a constant width.
Aside from the $M_{1-}^{3/2}$ multipole,
the energy dependence of the width becomes important
to account for fine details of some multipoles, 
such as the cusp peak that appears in the $E^p_{0+}$
electromagnetic multipole \cite{fernandez06a,SAID}, which is due to
the opening of the $\eta$ decay channel of the N(1535) resonance.

Let us now discuss the results obtained with the models I to V.
In order to treat each model on its own foot we have fitted 
the parameters to the data independently for each model. 
We fit the calculated electromagnetic multipoles to data
provided by the energy-independent solution of SAID \cite{SAID},
up to spin-3/2 and up to 1.2 GeV photon energy in the laboratory frame,
using masses and widths from \cite{Vrana}.
We use the optimization technique described in \cite{fernandez06a} .
For further details on the fitting procedure
we refer the reader to \cite{fernandez06a,fernandez06c}.
In these fits, the intrinsic E2/M1 ratio (EMR) of the $\Delta$(1232) \cite{fernandez06b}
is an output of the fit. In all the fits it is
consistent with the latest results from Lattice QCD \cite{Alexandrou05} within the error bars, that is
EMR$=\left( -1.93 \pm 0.94 \right)$\% for $Q^2=0.1$ GeV$^2$ and $m_\pi = 0$; and
EMR$=\left( -1.40 \pm 0.60 \right)$\% for $Q^2=0.1$ GeV$^2$ and $m_\pi = 370$ MeV.
To summarize, the five models considered are:
\begin{itemize}
\item[I]: Eq. (\ref{eq:width2}), $s$ and $u$ channels with
$\Gamma=\Gamma (s,u)$  (model in Ref. \cite{fernandez06a});
\item[II]: Eq. (\ref{eq:width4}), $s$ and $u$ channels with
constant width $\Gamma=\Gamma_0$;
\item[III]: Eq. (\ref{eq:width3}), 
only $s$ channel with constant width $\Gamma=\Gamma_0$;
\item[IV]: Eq. (\ref{eq:width1}),  
only $s$ channel with $\Gamma=\Gamma (s)$;
\item[V]: Eq. (\ref{eq:garcilazo}), $s$ channel with
$\Gamma=\Gamma(s)$ and $u$ channel with $\Gamma=0$.
\end{itemize}

Appart from the treatment of the resonance crossed terms and 
widths, the five models are constructed in the same way.
FSI are included through the inclusion of a phase to the electromagnetic multipoles
which matches the total phase as discussed in Ref. 
\cite{fernandez06a,fernandez06b,fernandez06c}
and, in particular, the same spin-3/2 couplings are also used in models
I to V. Expressions for the Lagrangians and
electromagnetic multipoles can be found in the same references.
As remarked in \cite{fernandez06a} the choice of the spin-3/2
couplings is very important.
For many years it has been customary to choose 
for the spin-3/2 Lagrangians the coupling scheme of Ref. 
\cite{Nath71} that presents pathologies
such as \cite{fernandez06a,Pascalutsa98}: 
spin 1/2 pollution, quantization anomalies, 
non-positive definite commutators, acausal fields,
as well as  bad threshold and high energy behaviors. 
In that scheme, the $u$ channel provides a too large contribution 
in the high energy region. To regularize their contribution
one has to include an extra cutoff in the crossed terms \cite{Garcilazo93},
which explicitly breaks crossing symmetry.
In our calculations, to avoid all these problems we use the spin-3/2 
coupling scheme suggested by Pascalutsa \cite{Pascalutsa98} 
that avoids all these pathologies and provides amplitudes that
behave properly in both the low and high-energy regions
\cite{fernandez06a}.

The energy-dependent widths have
been parametrized as in Ref. \cite{fernandez06a}
so that they fulfill the following physical requirements:
\begin{itemize}
\item[(a)] $\Gamma=\Gamma_0$ at $\sqrt{s}=M^*$; 
\item[(b)] $\Gamma \to 0$ when $k_\pi \to 0$, where $k$ is the
three-momentum of the outgoing pion in the center of mass reference system;
\item[(c)] $\Gamma$ has the 
correct angular momentum barrier at threshold, $k_\pi^{2L+1}$, with
$L$ the angular momentum of the resonance.
$X_j\left( l \right)$ in Eq. (\ref{eq:Xj}) is given by
\begin{equation}
X_j \left( l \right) = 2 \frac{\left(\frac{ k_j    }{  k_{j0} } 
\right)^{2L+1}}{1+\left( \frac{ k_j  }{  k_{j0} }
\right)^{2L+3}} \: \Theta \left( l - \left( M + m_j \right)^2 \right) ,
\end{equation}
where $M$ is the mass of the nucleon, $m_j$ stands for the mass of the meson of the  corresponding decay channel
$j = \pi , \pi \pi , \eta$, and
\begin{equation}
k_j=\sqrt{\left(l-M^2-m_j^2 \right)^2-4m_j^2M^2}/
\left( 2 \sqrt{l}\right) ,
\end{equation}
with $k_{j0} = k_j$ 
when $l=M^{*2}$.
\end{itemize}

\begin{table}
\caption{Comparison of $\chi^2$ values obtained with the different
choices for widths and crossed terms.} \label{tab:chi2}
\begin{tabular}{llc}

\hline

Model&Eq. & $\chi^2/\chi^2_{\text{Model I}}$\\

\hline

\textbf{I}: $\Gamma=\Gamma (s,u)$, model in Ref. \cite{fernandez06a} 
&(\ref{eq:width2}) & 1\\
\textbf{II}: $\Gamma=\Gamma_0$, constant width
&(\ref{eq:width4}) & 1.35\\
\textbf{III}: $\Gamma=\Gamma_0$, constant width
&(\ref{eq:width3}) & 1.85\\
\textbf{IV}: $\Gamma=\Gamma (s)$, $s$ channel
&(\ref{eq:width1}) & 1.51\\
\textbf{V}: $\Gamma=\Gamma (s)$, traditional width scheme 
&(\ref{eq:garcilazo}) & 1.18\\

\hline

\end{tabular}
\end{table}

In Table \ref{tab:chi2} we compare the $\chi^2$ obtained from the fits with
models I to V.

Notably, models that take into 
account $u$ channels provide the best $\chi^2$ (models I, II, and V), with
better $\chi^2$ for those which include an energy dependence in the widths (I and V).
This is due to the large energy range covered by all the models, where a constant width is less reliable.
Between the two remaining fits, the best fit is obtained by the model which exhibits crossing symmetry.
The other models (III and IV) provide $\chi^2$ more than a 50\% larger, and have
the same number of parameters. Hence, it can be concluded that
$u$ channels make a difference when it comes to describe the experimental data.

\begin{figure}
\begin{center}
\rotatebox{0}{\scalebox{0.4}[0.4]{\includegraphics{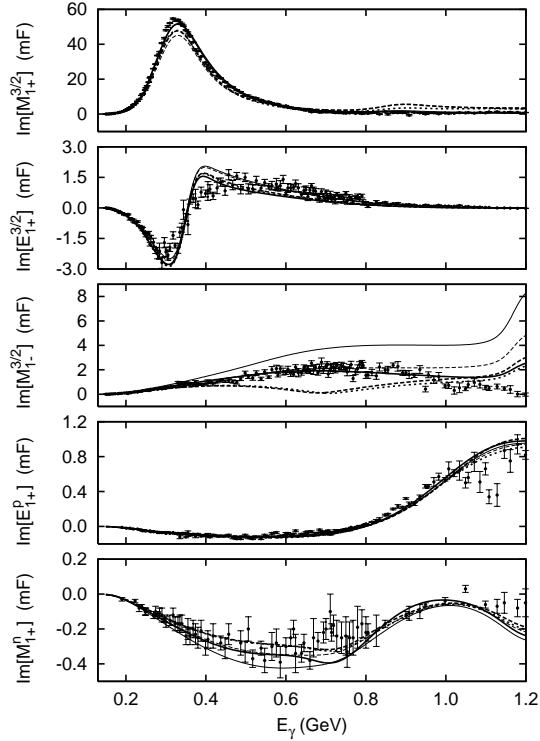}}}
\end{center}
\caption{Complete calculation of the electromagnetic
multipoles presented in Fig. \ref{fig:fig-crossing}
(FSI included) compared to data from \cite{SAID}.
Each model has been fitted to data independently, 
corresponding $\chi^2$ values are
provided in Table \ref{tab:chi2}.
Curve conventions: 
Thick solid:  Model I, Eq. (\ref{eq:width2}); 
Thick dashed: Model II, Eq. (\ref{eq:width4}) (constant width); 
Thin dashed:  Model III, Eq. (\ref{eq:width3}) 
(constant width, no $u$ channel);
Thin solid:   Model IV, Eq. (\ref{eq:width1}); 
Short dashed: Model V, Eq. (\ref{eq:garcilazo}) 
(traditional width scheme).} 
\label{fig:fig-crossing2}
\end{figure}

In Fig. \ref{fig:fig-crossing2} we show the comparison to the data \cite{SAID}
of our results for a few electromagnetic multipoles 
(Im$\left[  M_{1+}^{3/2} \right]$, Im$\left[ E_{1+}^{3/2} \right]$,
Im$\left[ M_{1-}^{3/2} \right]$, Im$\left[ E_{1+}^p \right]$, and Im$\left[ M_{1+}^n \right]$)
including FSI.
For the well-established resonant $M^{3/2}_{1+}$ and $E^{3/2}_{1+}$ multipoles of the $\Delta$(1232)
all the models provide similar results (upper panels in Fig. \ref{fig:fig-crossing2}).
For most  multipoles, the overall behavior is reproduced with
the constant-width model (model II) that also
respects crossing symmetry, but the $\chi^2$ is smaller when energy dependent widths are
considered (model I).
As observed in Figs \ref{fig:fig-crossing} and \ref{fig:fig-crossing2} 
the multipole Im$\left[ M_{1-}^{3/2} \right]$,
is more sensitive to the choice of the width, which only is well described by models I and IV (crossing symmetric with energy dependent widths).
In the higher energy region ($E_\gamma > 1$ GeV ),
the description of data on Im$\left[ M_{1-}^{3/2} \right]$ multipole
is not satisfactory for any model.
This shows that high-lying resonances may not be well accounted for.
Three star resonances, which have not been included, may play a role
in the improvement of the data description in the high-energy region.

\begin{figure}
\begin{center}
\rotatebox{0}{\scalebox{0.4}[0.4]{\includegraphics{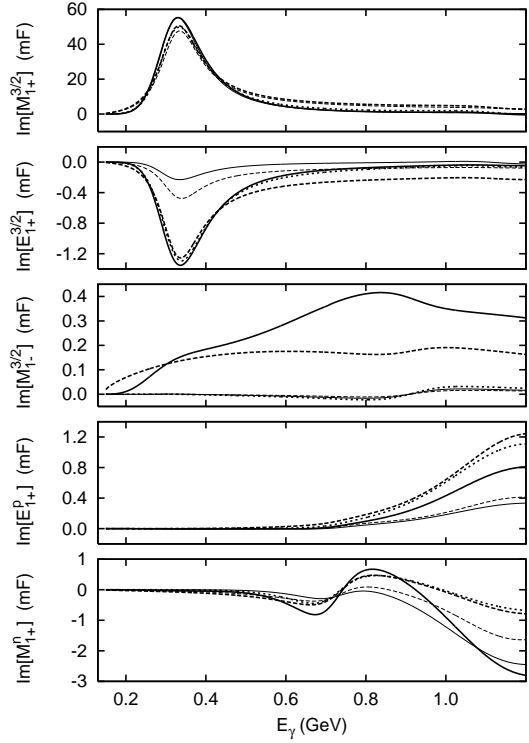}}}
\end{center}
\caption{Bare electromagnetic multipoles obtained with models I to V using their corresponding
coupling constants. Curve conventions as in Fig. \ref{fig:fig-crossing2}.} \label{fig:fig-crossing3}
\end{figure}

In Fig. \ref{fig:fig-crossing3} we compare the bare electromagnetic multipoles obtained 
using models I to V and their corresponding coupling constants.
For each multipole the results obtained with models I and II are similar.
However, the latter differ substantially from the other models III, IV, and V.
The comparison of the bare multipoles in Fig. \ref{fig:fig-crossing3} to their corresponding 
dressed multipoles in Fig. \ref{fig:fig-crossing2} show that FSI play an important role
in neutral pion production and are essential to describe properly the imaginary part of the
electromagnetic multipoles.

We conclude that the inclusion of the width in the $u$ channel
in a crossing symmetric way is not merely academic
but makes a significant difference as it stems from results in Table \ref{tab:chi2}.
Certain observables such as the  $M_{1-}^{3/2}$ and $M_{1+}^n$ multipoles are
particularly sensitive to the $u$ channel contribution.
The influence of the imaginary part of the resonance amplitude is
not important in most of the multipoles, but it makes a difference in the imaginary parts
of the $M_{1-}^{3/2}$ and $M_{1+}^n$.
Actually, in $\text{Im}M_{1-}^{3/2}$ multipole, the $u$ channels of the resonances
play a more important role than the direct
channel contributions, which are dominated by the $\Delta$(1620).
This multipole is highly interesting from the theoretical pointy
of view as it offers the possibility
to study the effects of crossing symmetry
and the energy dependence of the widths.
It will be very interesting to test the different models also in pion photoproduction from nuclei
where the bare amplitudes should, in principle, be used.

\begin{ack}
This work has been supported in part under contracts
FIS2005-00640, FPA2006-07393, and FPA2007-62216
of Ministerio de Educaci\'on y Ciencia (Spain) and by UCM and Comunidad de Madrid
under project number 910059 (Grupo de F\'{\i}sica Nuclear).
The computations of this work were carried out at the 
``Cluster de C\'alculo de Alta Capacidad para T\'ecnicas F\'{\i}sicas''
partly funded by EU Commission under FEDER programme and by 
Universidad Complutense de Madrid (Spain).
\end{ack}

\end{document}